\documentclass[a4paper,12pt]{article}%
\usepackage{eurosym}
\usepackage{amsfonts}
\usepackage{amssymb}
\usepackage{amsmath,amsthm}
\usepackage{graphicx}
\usepackage{hyperref}
\usepackage{accents}
\usepackage{xcolor}
\usepackage{float,rotating,subfigure}
\usepackage{longtable,lscape}
\usepackage{blindtext,pdflscape}
\usepackage[authoryear,round]{natbib}
\usepackage{amsmath}
\usepackage{pdflscape}
\usepackage{rotating}%
\providecommand{\U}[1]{\protect\rule{.1in}{.1in}}
\newtheorem{theorem}{Theorem}

\newtheorem{conjecture}[theorem]{Conjecture}
\newtheorem{corollary}{Corollary}

\newtheorem{lemma}{Lemma}

\newtheoremstyle{style2}{4pt}{4pt}{}{}{\bfseries}{.}{ }{}
\theoremstyle{style2}

\hypersetup{colorlinks=true,
linkcolor=blue,
citecolor=blue,
urlcolor=blue}
\begin{document}

\title{On the Robustness of Mixture Models in the Presence of Hidden Markov Regimes
with Covariate-Dependent Transition Probabilities\thanks{The authors wish to
thank Patrik Guggenberger, Peter Phillips and two referees for their helpful
comments and suggestions. For the purposes of open access, the corresponding
author has applied a CC-BY public copyright licence to any accepted manuscript
version arising from this submission. Address correspondence to: Zacharias
Psaradakis, Birkbeck Business School, Birkbeck, University of London, Malet
Street, London WC1E 7HX, UK; e-mail: z.psaradakis@bbk.ac.uk. }}
\author{\textbf{Demian Pouzo}\\{\normalsize Department of Economics, University of California, Berkeley,
U.S.A.}\\{\normalsize Email:} {\normalsize dpouzo@berkeley.edu}
\and \textbf{Zacharias Psaradakis}\\{\normalsize Birkbeck Business School, Birkbeck, University of London, U.K.}\\{\normalsize Email:} {\normalsize z.psaradakis@bbk.ac.uk}
\and \textbf{Martin Sola}\\{\normalsize Department of Economics, Universidad Torcuato di Tella,
Argentina}\\{\normalsize Email:} {\normalsize msola@utdt.edu}}
\date{}
\maketitle

\begin{abstract}
\noindent This paper studies the robustness of quasi-maximum-likelihood (QML) estimation in hidden Markov models (HMMs) when the regime-switching structure is misspecified. Specifically, we examine the case where the true data-generating process features a hidden Markov regime sequence with covariate-dependent transition probabilities, but estimation proceeds under a simplified mixture model that assumes regimes are independent and identically distributed. We show that the parameters governing the conditional distribution of the observables can still be consistently estimated under this misspecification, provided certain regularity conditions hold. Our results highlight a practical benefit of using computationally simpler mixture models in settings where regime dependence is complex or difficult to model directly.
\bigskip

\noindent\noindent\textit{Key words and phrases}: Consistency;
covariate-dependent transition probabilities; identifiability; hidden Markov
model; mixture model; quasi-maximum-likelihood; misspecified model.\bigskip
\newpage

\end{abstract}

\baselineskip0.3in

\section{Introduction}

\label{Intro}Consistency and asymptotic normality of least-squares estimators
in regression models in the presence of potential model misspecification ---
e.g., misspecification of the response function or misspecification of the
dynamic structure of the errors --- are well-established facts (see, e.g.,
\cite{DomowitzWhite82}). Such fundamental results, together with the related
classical work of \cite{Huber67}, underpin a large body of literature
exploring the feasibility of drawing valid and meaningful inferences from
parametric models that need not necessarily contain the true data-generating
process (DGP). Numerous results of this kind have been established for a wide
variety of models and estimators, both in static and dynamic settings, ranging
from inference procedures based on estimating equations and moment conditions
(e.g., \cite{Bates85}) to quasi-maximum-likelihood (QML) procedures for
conditional mean, conditional variance and conditional quantile models (e.g.,
\cite{White82,white94}, \cite{Levine83}, \cite{Gourieroux84},
\cite{NeweySteigerwald97}, \cite{KOMUNJER2005}).

This paper adds to the literature by presenting another example of robustness
with respect to misspecification. Specifically, we consider the case of Hidden
Markov Models (HMMs), where observable variables exhibit conditional
independence given an underlying unobservable regime sequence (and, possibly,
exogenous covariate sequences), focusing on situations where the dependence
structure of the regime sequence is misspecified. In our set-up, the DGP is
taken to be a generalized HMM that may include covariates and has a finite
number of Markov regimes, but the postulated probability model is a finite
mixture model, that is, an HMM with independent, identically distributed
(i.i.d.) regimes. By considering the pseudo-true parameter set for the QML
estimator in the (misspecified) mixture model, it is shown that the parameters
of the conditional distribution of the observable response variables are
consistently estimable even if the dependence of the unobservable regime
sequence is not taken into account. A condition on the tail behavior of the
characteristic function of the (standardized) conditional distribution of the
observable responses is also provided under which the pseudo-true parameter
for the QML estimator is a singleton set. An important distinguishing feature
of our analysis is that the true regime sequence is allowed to be a temporally
inhomogeneous Markov chain whose transition probabilities are functions of
observable variables.

This case holds practical significance given the widespread use of both HMMs
and mixture models. HMMs with temporally inhomogeneous regime sequences have
found applications in diverse areas such as biology (e.g.,
\cite{ghavidel2015nonhomogeneous}), economics (e.g., \cite{dieb94},
\cite{Engel96}), earth sciences (e.g., \cite{Hughes99}), and engineering
(e.g., \cite{ramesh1992modeling}). Temporally homogenous variants of HMMs and
of Markov-switching regression models are also used extensively in economics
and finance (e.g., \cite{engelhamilton90}, \cite{ryden98}, \cite{Jeanne2000},
\cite{bollen0}), as well as in biology, computing, engineering and statistics
(see \cite{Ephraim2002} and references therein). Statistical inference in such
models is typically likelihood-based and the properties of QML procedures are,
naturally, of much interest. Nevertheless, HMMs are inherently intricate and
computationally demanding due to the need to account for the underlying
correlated regime sequence and for the dependence of the conditional
distribution on the current hidden regime. By demonstrating that it is
feasible to use a mixture model ---~a simpler and computationally less
demanding framework~--- while still estimating consistently the parameters of
the conditional distribution of the observations, this paper offers a more
accessible avenue for practitioners to follow without sacrificing the accuracy
of parameter estimates.

In related recent work, \citet{pouzo22MLE} considered the asymptotic
properties of the QML estimator in a rich class of models with Markov regimes
under general conditions which allow for autoregressive dynamics in the
observation sequence, covariate-dependence in the transition probabilities of
the hidden regime sequence, and potential model misspecification. The QML
estimator was shown to be consistent for the pseudo-true parameter (set) that
minimizes the Kullback--Leibler information measure. Unsurprisingly,
identifying the possible limit of the QML estimator when the true probability
structure of the data does not necessarily lie within the parametric family of
distributions specified by the model is not always a feasible task in such a
general set-up. This paper provides an answer in the simpler case of
switching-regression models, HMMs and related mixture models. Consistency
results for misspecified pure HMMs (with no covariates in the outcome
equation) can also be found in \cite{Mevel04} and \cite{douc12}. Unlike our
analysis, which allows the regime transition probabilities to be
time-dependent and driven by observable variables, these papers restrict
attention to the case of time-invariant transition mechanisms.

In the next section, we introduce the DGP and statistical model of interest,
and consider QML estimation of the parameters of the outcome equation of a
misspecified generalized HMM. Section~\ref{MonteCarlo} discusses numerical
results from a simulation study. Section~\ref{conclusion} summarizes and concludes.

\section{Framework, Results and Discussion}

\label{mixture}

\subsection{DGP and Model}

Consider a discrete-time stochastic process $\{(X_{t},S_{t})\}_{t\geq0}$ such
that $X_{t}=(Y_{t},Z_{t},W_{t})$ is an observable variable with values in
$\mathbb{X}\subset\mathbb{R}^{3}$ and $S_{t}$ is a latent variable with values
in $\mathbb{S}:=\{1,2,\ldots,d\}\subset\mathbb{N}$ for some $d\geq2$. The
variable $S_{t}$ is viewed as the hidden regime (or state) associated with
index $t$, which is \textquotedblleft observable\textquotedblright\ only
indirectly through its effect on $X_{t}$. The following assumptions are made
about the DGP:

\begin{enumerate}
\item For each $t\geq1$,~the conditional distribution of $S_{t}$ given
$X_{0}^{t-1}:=(X_{0},\ldots,X_{t-1})$ and $S_{0}^{t-1}:=(S_{0},\ldots
,S_{t-1})$, denoted by $Q_{\ast}(\cdot|Z_{t-1},S_{t-1})$, depends only on
$(Z_{t-1},S_{t-1})$ and is such that $Q_{\ast}(s|z,s^{\prime})>0$ for all
$(s,s^{\prime},z)\in\mathbb{S}^{2}\times\mathcal{Z}$, where $\mathcal{Z}%
\subset\mathbb{R}$ is the state space of $Z_{t}$.

\item For each $t\geq1$, the conditional distribution of $Z_{t}$ given
$(X_{0}^{t-1},S_{0}^{t})$ depends only on $Z_{t-1}$; furthermore,
$\{(Z_{t},S_{t})\}_{t\geq0}$ is strictly stationary with invariant
distribution $\nu_{ZS}$.

\item For each $t\geq1$,~the conditional distribution of $Y_{t}$ given
$(X_{0}^{t-1},S_{0}^{t},W_{t})$ depends only on $(W_{t},S_{t})$ and is
specified via the equation
\begin{equation}
Y_{t}=\mu_{1}^{\ast}(S_{t})+\gamma^{\ast}(S_{t})W_{t}+\sigma_{1}^{\ast}%
(S_{t})U_{1,t}, \label{hmm1}%
\end{equation}
where $\mu_{1}^{\ast}$, $\gamma^{\ast}$ and $\sigma_{1}^{\ast}>0$ are known
real functions on $\mathbb{S}$; the noise variables $\{U_{1,t}\}_{t\geq0}$ are
i.i.d., independent of $\{S_{t}\}_{t\geq0}$, with mean zero, variance one, and
density $f$.

\item For each $t\geq1$, the conditional distribution of $W_{t}$ given
$(X_{0}^{t-1},S_{0}^{t})$ depends only on $W_{t-1}$, and $W_{t}$ is
independent of $(Z_{t-1},S_{t-1})$; furthermore, $\{W_{t}\}_{\geq0}$ is
strictly exogenous in (\ref{hmm1}) and strictly stationary with invariant
distribution $\nu_{W}$. \emph{ }
\end{enumerate}

\medskip

Instead of the Markov-switching structure of the DGP, the researcher's
postulated parametric model is a family of finite mixture models (without
Markov dependence). Specifically, the model is specified by assuming that the
regime variables $\{S_{t}\}_{t\geq1}$ are i.i.d. with distribution
\begin{equation}
Q_{\bar{\vartheta}}(s)=\bar{\vartheta}_{s}\in(0,1),\quad s\in\mathbb{S}.
\label{mix1}%
\end{equation}
In addition, the observable variables $\{Y_{t}\}_{t\geq1}$ are assumed to
satisfy the equations
\begin{equation}
Y_{t}=\mu(S_{t})+\gamma(S_{t})W_{t}+\sigma(S_{t})\varepsilon_{t},\quad t\geq1,
\label{mix2}%
\end{equation}
where $\mu$, $\gamma$ and $\sigma>0$ are known real functions on $\mathbb{S}$
and $\{\varepsilon_{t}\}_{t\geq1}$ are i.i.d. random variables, independent of
$\{(S_{t},W_{t})\}_{t\geq1}$, such that $\varepsilon_{1}$ has the same density
$f$ as $U_{1,1}$. The mixture model defined by (\ref{mix1}) and (\ref{mix2})
is parameterized by $\theta:=(\pi(s),\bar{\vartheta}_{s})_{s\in\mathbb{S}}$,
with $\pi(s):=(\mu(s),\gamma(s),\sigma(s))$, which is assumed to take values
in a compact set $\Theta\subset\mathbb{R}^{q}$, $q>1$. We denote by $P_{\pi
}(\cdot|W_{t},S_{t})$ the conditional distribution of $Y_{t}$ given
$(W_{t},S_{t})$ that is implied by (\ref{mix2}); the corresponding conditional
density is denoted by $p_{\pi}(\cdot|W_{t},S_{t})$.

\paragraph{Key aspects of our set-up:}

First, the DGP has a (generalized) HMM structure in which $\{Y_{t}\}_{t\geq0}$
are independent, conditionally on the regime sequence $\{S_{t}\}_{t\geq0}$ and
an exogenous covariate sequence $\{W_{t}\}_{t\geq0}$ (having the Markov
property), so that the conditional distribution of $Y_{t}$ given the regime
and covariate sequences depends only on $(S_{t},W_{t})$. The inclusion of the
exogenous covariate $W_{t}$ in (\ref{hmm1}) and (\ref{mix2}) allows the study
of the causal effect of $W$ on $Y$ under different regimes; this causal effect
is captured by $\gamma^{\ast}$ and is estimable via the mixture specification
(\ref{mix1})--(\ref{mix2}). Exogeneity of $W$ (assumption 4 above) is
essential for the results discussed in Section~\ref{qml}~to hold and, hence,
for consistent estimation of the causal effect $\gamma^{\ast}$ under the
(erroneous) assumption of independent regimes.\footnote{The standard HMM
formulation is a special case in which $W_{t}$ is absent from the outcome
equation (\ref{hmm1}).} Second, the true hidden regimes $\{S_{t}\}_{t\geq0}$
are a temporally inhomogeneous Markov chain whose transition probabilities
depend on the lagged value of the observable variable $Z_{t}$. The sequence
$\{Z_{t}\}_{t\geq0}$ has the Markov property and is not required to be
exogenous, in the sense that $Z_{t}$ may be contemporaneously correlated with
$U_{1,t}$. Third, the statistical model is misspecified, in the sense that the
DGP is not a member of the family $\{(P_{\pi},Q_{\bar{\vartheta}})\colon
(\pi,\bar{\vartheta})\in\Theta\}$; this is because the dynamic structure of
the regimes is misspecified. As already discussed in Section~\ref{Intro}, this
relatively simple set-up is of much practical interest since HMMs with
temporally inhomogeneous regime sequences have found many applications.
Mixture models with i.i.d. regimes are also widely used in many different
fields (see \cite{McLachlan2000} and \cite{Schnatter06}), including economics
and econometrics (see \cite{Compiani2016}).

It is worth noting that, although we focus on scalar responses and covariates
for the sake of simplicity, all our results can be extended straightforwardly
to cases where $X_{t}\in\mathbb{X}\subset\mathbb{R}^{h}$ with $h>3$. For
example, $W_{t}$ may be a vector of covariates, which may include lagged
values of $W_{t}$ in cases where dynamic causal effects are of interest.
Similarly, $Z_{t}$ may be a vector of information variables that affect the
dynamic profile of the regime transition probabilities, whose generating
mechanism has a finite-order autoregressive structure.

\subsection{QML Estimation}

\label{qml}

Given observations $(X_{1},\ldots,X_{T})$, $T\geq1$, the quasi-log-likelihood
function for the parameter $\theta$ is%
\begin{equation}
\theta\mapsto\ell_{T}(\theta):=T^{-1}\sum_{t=1}^{T}\ln\left(  \sum
_{s\in\mathbb{S}}\bar{\vartheta}_{s}p_{\pi}(Y_{t}|W_{t},s)\right)  .
\label{likf}%
\end{equation}
The QML estimator $\hat{\theta}_{T}$ of $\theta$ is defined as an approximate
maximizer of $\ell_{T}(\theta)$ over $\Theta$, so that%
\[
\ell_{T}(\hat{\theta}_{T})\geq\sup_{\theta\in\Theta}\ell_{T}(\theta)-\eta
_{T},
\]
for some sequence $\{\eta_{T}\}_{T\geq1}\subset\mathbb{R}_{+}$ converging to zero.

It is not too onerous to verify that, under assumptions that are common in the
literature (e.g., Gaussianity of $U_{1,1}$ and $Q_{\ast}(s|z,s^{\prime
})=G(\alpha_{s,s^{\prime}}+\beta_{s,s^{\prime}}z)$ for some continuous
distribution function $G$ on $\mathbb{R}$ whose support is all of $\mathbb{R}%
$), the conditions of \citet{pouzo22MLE} required for convergence of the QML
estimator of $\theta$ to a well-defined limit are satisfied. Specifically, let%
\[
\theta\mapsto H^{\ast}(\theta):=\mathbb{E}_{\bar{P}_{\ast}}\left[  \ln\left(
\frac{p_{\ast}(Y_{1}|W_{1})}{p_{\theta}(Y_{1}|W_{1})}\right)  \right]
\]
be the Kullback--Leibler information function, where $p_{\theta}(Y_{1}%
|W_{1}):=\sum_{s\in\mathbb{S}}\bar{\vartheta}_{s}p_{\pi}(Y_{1}|W_{1},s)$
denotes the conditional density of $Y_{1}$ given $W_{1}$ induced by $(P_{\pi
},Q_{\bar{\vartheta}})$ for each $(\pi,\bar{\vartheta})\in\Theta$, $p_{\ast
}(Y_{1}|W_{1})$ denotes the conditional density of $Y_{1}$ given $W_{1}$
induced by the (true) DGP, and the expectation $\mathbb{E}_{\bar{P}_{\ast}%
}(\cdot)$ is with respect to the distribution $\bar{P}_{\ast}$ of
$\{(X_{t},S_{t})\}_{t\geq0}$ induced by the (true) DGP. Then, we have
\begin{equation}
\inf_{\theta\in\Theta_{\ast}}||\hat{\theta}_{T}-\theta||\rightarrow
0\text{\quad as }T\rightarrow\infty,\label{cons1}%
\end{equation}
in $\bar{P}_{\ast}$-probability, where
\begin{equation}
\Theta_{\ast}:=\underset{\theta\in\Theta}{\arg\min}~H^{\ast}(\theta
)\label{cons2}%
\end{equation}
is the pseudo-true parameter (set) and $\left\Vert \cdot\right\Vert $ denotes
the Euclidean norm on $\mathbb{R}^{q}$ (cf. Theorem~1 of \citet{pouzo22MLE}).

A sharper result can be established by considering the pseudo-true parameter
$\Theta_{\ast}$ under the specified DGP. Together with (\ref{cons1}) and
(\ref{cons2}), the following theorem shows that, despite the erroneous
treatment of hidden regimes as independent, QML based on the (misspecified)
mixture model provides consistent estimators of the true parameters of the
outcome equation.

\begin{theorem}
\label{th1}The choice $\mu=\mu_{1}^{\ast}$, $\sigma=\sigma_{1}^{\ast}$,
$\gamma=\gamma^{\ast}$, and $(\bar{\vartheta}_{s}^{\ast})_{s\in\mathbb{S}}$
such that $\bar{\vartheta}_{s}^{\ast}=\mathbb{E}_{\nu_{ZS}}[Q_{\ast}(s|Z,S)]$
for all $s\in\mathbb{S}$ is a pseudo-true parameter, that is, it maximizes the
function
\[
\theta\mapsto\mathbb{E}_{\bar{P}_{\ast}}\left[  \ln\left(  \sum_{s\in
\mathbb{S}}\frac{\bar{\vartheta}_{s}}{\sigma(s)}f\left(  \frac{Y_{1}%
-\mu(s)-\gamma(s)W_{1}}{\sigma(s)}\right)  \right)  \right]  .
\]

\end{theorem}

\begin{proof}
	Observe that the Kullback--Leibler information function $H^{\ast}$ is proportional to
	\begin{align}\label{eqn:mixture.id.1}
	\theta\mapsto-\int_{\mathbb{R}^{2}}\ln\left(  \frac{\sum_{s\in\mathbb{S}}%
	\bar{\vartheta}_{s} \sigma(s)^{-1}f\left(  (y-\mu(s) - \gamma(s)w)/\sigma(s)\right)
	}{p_{\ast}(y|w)}\right)  p_{\ast}(y|w)  dy\nu_{W}(dw),
	\end{align}
	where
	\[
	(y,w) \mapsto p_{\ast}(y|w)  = \sum_{s\in\mathbb{S}}\mathrm{Pr}_{\ast}(S_{1}
	=s \mid W_{1} = w)\sigma_{1}^{\ast}(s)^{-1}f\left(  (y-\mu_{1}^{\ast}(s) - \gamma^{\ast}(s) w  )/\sigma_{1}^{\ast
	}(s)\right).
	\]
	Under our assumptions about $(W_{t},Z_{t-1})$, $\mathrm{Pr}_{\ast}(S_{1}=\cdot | W_{1} = w) = \mathrm{Pr}_{\ast}(S_{1}=\cdot)$, where $\mathrm{Pr}_{\ast}$ stands for the true probability over the hidden
	regimes, given by
	\[
	s\mapsto\mathrm{Pr}_{\ast}(S_{1}=s) : =\int_{\mathbb{R}\times\mathbb{S}}%
	\sum_{s^{\prime}\in\mathbb{S}}Q_{\ast}(s|z,s^{\prime})\nu_{ZS}(dz,ds^{\prime
	}).
	\]
The minimizers of the function in (\ref{eqn:mixture.id.1}) are all $\theta$ such
	that
	\[
	\sum_{s\in\mathbb{S}}\bar{\vartheta}_{s} \sigma(s)^{-1}f\left(  (\cdot
	-\mu(s) - \gamma(s) \cdot )/\sigma(s)\right)  =p_{\ast}(\cdot | \cdot).
	\]
	It is straightforward to verify that the equality above holds for $\mu=\mu
	_{1}^{\ast}$, $\sigma=\sigma_{1}^{\ast}$, $\gamma=\gamma^{\ast}$, and $\bar{\vartheta}^{\ast}$ such that
	$\bar{\vartheta}^{\ast}_{s}=\mathrm{Pr}_{\ast}(S_{1}=s)$.		
\end{proof}

Theorem~\ref{th1} establishes that the true parameters $\pi^{\ast}%
(s):=(\mu_{1}^{\ast}(s),\sigma_{1}^{\ast}(s),\gamma^{\ast}(s))$,
$s\in\mathbb{S}$, associated with the observation equation (\ref{hmm1}),
together with $(\bar{\vartheta}_{s}^{\ast})_{s\in\mathbb{S}}$, minimize the
Kullback--Leibler information $H^{\ast}(\theta)$. The corollary that follows
shows that this minimizer is unique, as long as $\theta_{\ast}:=(\pi^{\ast
}(s),\bar{\vartheta}_{s}^{\ast})_{s\in\mathbb{S}}$ is identified in $\Theta$.
In the present context, $\theta_{\ast}$ is said to be identified in $\Theta$
if, for any $\theta\in\Theta$ such that $p_{\theta}(\cdot|\cdot)=p_{\theta
_{\ast}}(\cdot|\cdot)$ $\bar{P}_{\ast}$-almost surely, $\theta=\theta_{\ast}$
up to permutations.\footnote{Formally, $(\pi(\mathfrak{p}[s]),\bar{\vartheta
}_{\mathfrak{p}[s]})=(\pi^{\ast}(s),\bar{\vartheta}_{s}^{\ast})$ for all
$s\in\mathbb{S}$ and any permutation $\mathfrak{p}:\mathbb{S}\rightarrow
\mathbb{S}$. The qualifier `up to permutations' reflects the fact that the
problem remains unchanged if the indices of the regimes are permuted.}

\begin{corollary}
\label{cor:unique} If $\theta_{\ast}$ is identified in $\Theta$, then it is
the unique maximizer (up to permutations) of
\[
\theta\mapsto\mathbb{E}_{\bar{P}_{\ast}}\left[  \ln\left(  \sum_{s\in
\mathbb{S}}\frac{\bar{\vartheta}_{s}}{\sigma(s)}f\left(  \frac{Y_{1}%
-\mu(s)-\gamma(s)W_{1}}{\sigma(s)}\right)  \right)  \right]  .
\]

\end{corollary}

\begin{proof}
	By Theorem \ref{th1}, it suffices to show that 	\begin{align*}
	\int_{\mathbb{R}^{2}}\ln\left( p_{\ast}(y|w) \right)  p_{\ast}(y|w)  dy \nu_{W}(dw)
>	\int_{\mathbb{R}^{2}}\ln\left(  p_{\theta}(y|w)   \right)  p_{\ast}(y|w)  dy\nu_{W}(dw),
	\end{align*}
for any $\theta\in\Theta$ which is not equal to a permutation of $\theta_{\ast}$.
Since $\theta_{\ast}$ is identified, for any such $\theta$, $  p_{\theta}( \cdot | \cdot)   \ne   p_{\theta_{\ast}}( \cdot | \cdot)  $ with positive probability under $\bar{P}_{\ast}$. Therefore, the strict inequality above follows from the (strict) Jensen inequality.
\end{proof}

To provide some (non-technical) intuition behind Theorem \ref{th1} and its
corollary, recall that when the misspecified mixture model is fitted to data
using maximum likelihood techniques, the objective function which is maximized
is a quasi-likelihood. The resulting QML estimator is an estimator for the
parameters which make the model's implied conditional distribution of the
response as close as possible --- measured in Kullback--Leibler divergence ---
to the true conditional distribution, even when the model is misspecified.
Despite the fact that the researcher ignores dependence of the regimes, the
conditional distribution of the response variable given the covariates remains
a mixture of distributions under both the true and misspecified models. This
structural similarity allows the mixture model to match the key features of
the conditional distribution of interest correctly and, thus, the QML
estimator consistently recovers the parameters of the outcome equation. This
result relies heavily on two key conditions: stationarity of the regimes
(Assumption~2) and exogeneity of the covariates (Assumption~4). The former
stationarity condition is crucial because it ensures stability of the marginal
distribution of the regimes, which the mixture model tries to fit. Without
stationarity, the limiting object of the QML estimation procedure could vary
over time, and consistency would break down. The latter exogeneity condition
is fundamental because it guarantees that the covariates do not
\textquotedblleft carry\textquotedblright\ information about future states or
future shocks into the noise of the outcome equation. If exogeneity failed,
the conditional distribution of the response would not be properly captured by
the simple mixture model and the QML estimator would be biased and
inconsistent for the parameters of interest.

We conclude by remarking that when the minimizer $\theta_{\ast}$ of the
Kullback--Leibler information function is identified in $\Theta$ (and belongs
to the interior of $\Theta$), asymptotic normality of $\sqrt{T}(\hat{\theta
}_{T}-\theta_{\ast})$ may be deduced from the results of \citet{pouzo22MLE}
under suitable differentiability and moment conditions. These conditions are
satisfied, for example, in the case where $f$ is Gaussian and $Q_{\ast
}(s|z,s^{\prime})=G(\alpha_{s,s^{\prime}}+\beta_{s,s^{\prime}}z)$ for some
continuous distribution function $G$ on $\mathbb{R}$ whose support is all of
$\mathbb{R}$. In the next subsection, we discuss a sufficient condition for
the high-level identifiability requirement of Corollary~\ref{cor:unique} and
show that this condition holds in some commonly used models.

\subsection{Discussion on the Identifiability Condition}

Identifiability of mixture models has been studied extensively in the
literature following the original contribution of \cite{Teicher1963}, who
established identifiability of finite mixtures of distributions such as the
one-dimensional Gaussian and gamma. \cite{Yakowitz1968} gave a necessary and
sufficient condition for identifiability, which holds, for example, in the
case of finite mixtures of multivariate Gaussian distributions. This condition
was exploited by \cite{Holzmann2004} and \cite{Holzmann2006} to provide
sufficient low-level identifiability conditions based on the tail behavior of
the characteristic function of the component distributions.\footnote{A review
of related results for parametric and nonparameteric models that incorporate
mixture distributions can be found in \cite{Compiani2016}.}

Using the approach of \cite{Holzmann2004} and \cite{Holzmann2006}, we now
present a low-level condition, based only on features of $f$, which is
sufficient for the identifiability of $\theta_{\ast}$ required in
Corollary~\ref{cor:unique}.

\begin{lemma}
\label{lem:mixture.ID} Let $\varphi$ be the characteristic function associated
with $f$. If, for any $a_{1}>a_{2}$,
\[
\lim_{\tau\rightarrow\infty}\frac{\varphi(a_{1}\tau)}{\varphi(a_{2}\tau)}=0,
\]
then $\theta_{\ast}$ is identified in $\Theta$.
\end{lemma}

\begin{proof}
By the classical result of \cite{Yakowitz1968}, to establish unique
identifiability, it suffices to show that $\left\{ \frac{1}{\sigma (s)}%
f\left( \frac{y-m(s,w)}{\sigma (s)}\right) \right\} _{s\in \mathbb{S}}$ are
linearly independent, where $m(s,w):=\mu (s)+\gamma (s)w$. To this end,
observe that
\[
\int_{\mathbb{R}}e^{\mathrm{i}\tau y}\frac{1}{\sigma (s)}f\left( \frac{%
y-m(s,w)}{\sigma (s)}\right) dy=e^{\mathrm{i}\tau m(s,w)}\int_{\mathbb{R}}e^{%
\mathrm{i}\tau u\sigma (s)}f\left( u\right) du=e^{\mathrm{i}\tau
m(s,w)}\varphi (\tau \sigma (s)).
\]%
Hence, for any $\lambda _{1},\ldots ,\lambda _{d}$ in $\mathbb{R}$, $%
\sum_{s\in \mathbb{S}}\lambda _{s}\frac{1}{\sigma (s)}f\left( \frac{y-m(s,w)%
}{\sigma (s)}\right) =0$ implies
\begin{equation}
\sum_{s\in \mathbb{S}}\lambda _{s}e^{\mathrm{i}\tau m(s,w)}\varphi (\tau
\sigma (s))=0,  \label{lem1}
\end{equation}%
for any $\tau \in \mathbb{R}$ and any $w\in \mathbb{R}$. Without loss of
generality, let $\sigma (1)\leq \sigma (2)\leq \cdots \leq \sigma (d)$, with
$m\in \mathbb{S}$ being such that $\sigma (1)=\cdots =\sigma (m)<\sigma (m+1)
$. Then, (\ref{lem1}) is equivalent to
\begin{equation}
\lambda _{1}+\sum_{s=2}^{m}\lambda _{s}e^{\mathrm{i}\tau \lbrack
m(s,w)-m(1,w)]}+\sum_{s=m+1}^{d}\lambda _{s}e^{\mathrm{i}\tau \lbrack
m(s,w)-m(1,w)]}\frac{\varphi (\tau \sigma (s))}{\varphi (\tau \sigma (1))}=0.
\label{lem2}
\end{equation}%
Since $\sigma (s)>\sigma (1)$ for any $s\in \{m+1,\ldots ,d\}$ and $e^{%
\mathrm{i}\tau \lbrack m(s,w)-m(1,w)]}$ is uniformly bounded in $\tau $, it
follows by our assumption that $\sum_{s=m+1}^{d}\lambda _{s}e^{\mathrm{i}%
\tau \lbrack m(s,w)-m(1,w)]}\frac{\varphi (\tau \sigma (s))}{\varphi (\tau
\sigma (1))}\rightarrow 0$ as $\tau \rightarrow \infty $. This result readily implies that $n^{-1} \sum_{l=1}^{n} \sum_{s=m+1}^{d}\lambda _{s}e^{\mathrm{i}
l u_{0} \lbrack m(s,w)-m(1,w)]}\frac{\varphi (l u_{0} \sigma (s))}{\varphi (l u_{0}
\sigma (1))}\rightarrow 0$ as $n \rightarrow \infty $, for any $u_{0} >0$.
Regarding the term
$\sum_{s=2}^{m}\lambda _{s}e^{\mathrm{i}\tau \lbrack m(s,w)-m(1,w)]}$,
observe that $m(s,w)\neq m(1,w)$ for any $s\in \{1,\ldots ,m\}$ ---
otherwise, since $\sigma (s)=\sigma (1)$, the regimes associated with $%
S_{t}=s$ and $S_{t}=1$ would be identical rather than distinct. Thus, by
choosing $\tau =lu_{0}$, where $l\in\mathbb{N}$ and $u_{0}\in
\mathbb{R}_{+}$ is such that $u_{0}[m(s,w)-m(1,w)]\in (-\pi,\pi )$ for all $s\in
\{1,\ldots ,m\}$, it follows by Lemma 2.1 in \cite{Holzmann2004} that $%
n^{-1}\sum_{l=1}^{n}\sum_{s=2}^{m}\lambda _{s}e^{\mathrm{i}%
lu_{0}[m(s,w)-m(1,w)]}\rightarrow 0$ as $n\rightarrow \infty $. By these two
results and (\ref{lem2}),
\begin{eqnarray*}
\lambda _{1} &=&-\lim_{n\rightarrow \infty }n^{-1}\sum_{l=1}^{n}\left(
\sum_{s=2}^{m}\lambda _{s}e^{\mathrm{i}lu_{0}[m(s,w)-m(1,w)]}+%
\sum_{s=m+1}^{d}\lambda _{s}e^{\mathrm{i} l u_{0} \lbrack m(s,w)-m(1,w)]}\frac{%
\varphi (l u_{0} \sigma (s))}{\varphi (l u_{0} \sigma (1))}\right)  \\
&=&0.
\end{eqnarray*}%
By iterating on this procedure, it follows that $\lambda _{1}=\lambda
_{2}=\cdots =\lambda _{d}=0$, thereby establishing the desired result.
\end{proof}

As an example, consider what is, arguably, the most widely used class of
mixture models, namely those in which $f$ is Gaussian. In this case,
$\varphi(\tau)=e^{-\tau^{2}/2}$, $\tau\in\mathbb{R}$, and $\varphi(a_{1}%
\tau)/\varphi(a_{2}\tau)=e^{-(a_{1}^{2}-a_{2}^{2})\tau^{2}/2}$, $a_{1}>a_{2}$,
so the condition of Lemma~\ref{lem:mixture.ID} is satisfied. Thus, in the
Gaussian case, the QML estimator of the parameters of the mixture model
(\ref{mix1})--(\ref{mix2}) converges, in $\bar{P}_{\ast}$-probability, to
$\theta_{\ast}$. This result remains valid for non-Gaussian distributions,
including distributions with heavy tails (and finite variance). For instance,
the result holds if $f$ is the density of a (rescaled) Student-$t$
distribution with degrees of freedom $\upsilon>2$ (see Example~1 in
\cite{Holzmann2006}).

\subsection{Discussion on the Main Theorem}

\label{discuss}

The consistency results in (\ref{cons1})--(\ref{cons2}) and in
Theorem~\ref{th1} are quite general, in the sense that they cover misspecified
generalized HMMs with temporally inhomogeneous regime sequences and arbitrary
observation conditional densities. They imply that dependence of the regimes
in such HMMs may be safely ignored as long as the parameters of interest are
those of the conditional density of the observations given the regimes and the
covariates. It is important to note, however, that care should be taken in
estimating the asymptotic covariance matrix of the QML estimator since the
inverse of the observed information matrix is not necessarily a consistent
estimator in a misspecified model. Consistent estimation in this case
typically requires the use of an empirical sandwich estimator that does not
rely on the information matrix equality (cf. Theorem~5 of \citet{pouzo22MLE}).

Treating the regimes as an independent sequence simplifies likelihood-based
inference compared to the case of correlated Markov regimes. In the latter
case, an added difficulty, as demonstrated by \citet{pouzo22MLE}, is that
consistent QML estimation of the true parameter values in a model with Markov
regimes having covariate-dependent transition functions typically requires
joint analysis of equations such as (\ref{hmm1}) and the generating mechanism
of $\{Z_{t}\}$, even if the parameters of interest are only those associated
with (\ref{hmm1}). Furthermore, as pointed out by \cite{hamilton16}, rich
parameterizations of the transition mechanism of the regime sequence may not
necessarily be desirable when working with relatively short time series
because of legitimate concerns relating to potential over-fitting and
inaccurate statistical inference. In such cases, parsimonious specifications
which provide good approximations to key features of the data ---~and, in our
setting, consistent estimates of the parameters of interest~--- can be
attractive and useful.

Note that, for a class of regime-switching models in which the regime sequence
$\{S_{t}\}$ is a temporally homogeneous, two-state Markov chain, an
observation analogous to that implied by Theorem~\ref{th1} was made by
\cite{chowhite07}. They argued that the parameters of a model for the
conditional distribution of the observable variable $X_{t}$, given
$(X_{0}^{t-1},S_{0}^{t})$, can be consistently estimated by QML based on a
misspecified version of the model with i.i.d. regimes --- and exploited this
result to construct a quasi-likelihood-ratio test of the null hypothesis of a
single regime against the alternative hypothesis of two regimes. However,
\cite{Carter2012} demonstrated that consistency of the QML estimator for the
true parameters in such a setting does not, in fact, hold if the model and the
DGP contain an autoregressive component. This observation remains true in our
more general set-up with temporally inhomogeneous hidden regime sequences.
Specifically, a result analogous to that in Theorem~\ref{th1} does not hold
when lagged values of $Y_{t}$ are present as covariates in the outcome
equations (\ref{hmm1}) and (\ref{mix1}) (e.g., as in Markov-switching
autoregressive models). In this case, misspecification of the dependence
structure of the regimes will affect estimation of all the parameters, not
just those associated with the transition functions of the regime sequence.

\section{Numerical Examples}

\label{MonteCarlo}

As a numerical illustration of the results discussed in Section~\ref{mixture},
we report here findings from a small Monte Carlo simulation study in which the
effect on QML estimators of ignoring Markov dependence of hidden regimes is assessed.

In the experiments, artificial data are generated according to the generalized
HMM defined by (\ref{hmm1}), with regimes $\{S_{t}\}$ which form a Markov
chain on $\mathbb{S}=\{1,2\}$ such that%
\[
\Pr(S_{t}=s|S_{t-1}=s,Z_{t-1}=z)=[1+\exp(-\alpha_{s}^{\ast}-\beta_{s}^{\ast
}z)]^{-1},\quad s\in\{1,2\},\quad z\in\mathbb{R},
\]
and with $\{Z_{t}\}$ and $\{W_{t}\}$ satisfying the autoregressive equations
\[
Z_{t}=\mu_{2}^{\ast}+\psi^{\ast}Z_{t-1}+\sigma_{2}^{\ast}U_{2,t},
\]%
\[
W_{t}=\mu_{3}^{\ast}+\delta^{\ast}W_{t-1}+\sigma_{3}^{\ast}U_{3,t}.
\]
The noise variables $\{(U_{1,t},U_{2,t},U_{3,t})\}$ are i.i.d, Gaussian,
independent of $\{S_{t}\}$, with mean zero and covariance matrix
\[
\left[
\begin{array}
[c]{ccc}%
1 & \rho^{\ast} & \omega^{\ast}\\
\rho^{\ast} & 1 & 0\\
\omega^{\ast} & 0 & 1
\end{array}
\right]  .
\]
The parameter values are $\alpha_{1}^{\ast}=\alpha_{2}^{\ast}=2$, $\beta
_{1}^{\ast}=-\beta_{2}^{\ast}=0.5$, $\mu_{1}^{\ast}(1)=-\mu_{1}^{\ast}(2)=1$,
$\gamma^{\ast}(1)=0.5$, $\gamma^{\ast}(2)=1$, $\sigma_{1}^{\ast}(1)=\sigma
_{1}^{\ast}(2)=1$, $\mu_{2}^{\ast}=\mu_{3}^{\ast}=0.2$, $\psi^{\ast}%
=\delta^{\ast}=0.8$, $\sigma_{2}^{\ast}=\sigma_{3}^{\ast}=1$, and $\rho^{\ast
},\omega^{\ast}\in\{0,0.65\}$.

For each of 1000 samples of size $T\in\{200,800,1600,3200\}$ from this DGP,
estimates of the parameters of the outcome equation are obtained by maximizing
the quasi-log-likelihood function (\ref{likf}) associated with the mixture
model (\ref{mix1})--(\ref{mix2}), with $\Pr(S_{t}=1)=\bar{\vartheta}$ and
$\varepsilon_{t}\sim\mathcal{N}(0,1)$. Monte Carlo estimates of the bias of
the QML estimators of $\mu(1)$, $\mu(2)$, $\gamma(1)$, $\gamma(2)$,
$\sigma(1)$ and $\sigma(2)$ are reported in Table~\ref{MC-t1}. We also report
the ratio of the sampling standard deviation of the estimators to estimated
standard errors (averaged across replications for each design point). The
latter are computed using a sandwich\ estimator based on the Hessian and the
gradient of the quasi-log-likelihood function (cf.
\citet[Theorem 5]{pouzo22MLE}), with weights obtained from the Parzen kernel
and a data-dependent bandwidth selected by the plug-in method of
\cite{andrews91}.

The results for $\omega^{\ast}=0$ shown in the top panel of Table~\ref{MC-t1}
reveal that, although the estimators of $\mu(1)$ and $\mu(2)$ are somewhat
biased in the smallest of the sample sizes considered, finite-sample bias
becomes insignificant in the rest of the cases (regardless of the value of the
correlation parameter $\rho^{\ast}$), as is to be expected in light of the
result in Theorem~\ref{th1}. Furthermore, unless the sample size is small,
estimated standard errors are very accurate as approximations to the standard
deviation of the QML estimators.

The bottom panel of Table~\ref{MC-t1} contains results for a DGP with
$\omega^{\ast}=0.65$. A non-zero value for the correlation parameter
$\omega^{\ast}$ violates the exogeneity assumption about $W_{t}$ that is
maintained throughout Section~\ref{mixture} (and it is not obvious what the
limit point of the QML estimator based on (\ref{likf}) might be in this case).
The simulation results show that estimators of the parameters of the outcome
equation are significantly biased, even for the largest sample size considered
in the simulations. Biases in this case are clearly a consequence of the
mixture model being misspecified beyond the assumption of i.i.d. regimes, the
additional source of misspecification being the incorrect assumption of
uncorrelatedness of the covariate $W_{t}$ and the noise variable $U_{1,t}$.
The results relating to the accuracy of the estimated standard errors are not
substantially different from those obtained with $\omega^{\ast}=0$.

\begin{sidewaystable}[ptb]
\caption{Bias and Standard Deviation of QML Estimators (HMM)}
\label{MC-t1}
\vspace{1cm}
	\begin{tabular}
	[c]{llcccccclcccccc}\hline\hline
	$T$ &  & $\mu(1)$ & $\mu(2)$ & $\gamma(1)$ & $\gamma(2)$ & $\sigma(1)$ & $\sigma(2)$ &  & $\mu(1)$ &
	$\mu(2)$ &  $\gamma(1)$ & $\gamma(2)$ & $\sigma(1)$ & $\sigma(2)$\\\hline
	&  & \multicolumn{6}{c}{$\rho^{\ast
		}=0 $ $ ,  $ $\omega^{\ast}=0$} &  & \multicolumn{6}{c}{$\rho^{\ast
		}=0.65 $ $ ,  $ $\omega^{\ast}=0$}\\\cline{4-7}\cline{11-14}
	&  & \multicolumn{13}{c}{Bias}\\
	200  & & 0.093 & -0.022 & -0.033 & 0.018 & -0.125 & -0.045 & & 0.090 & -0.032 & -0.029 & 0.013 & -0.111 & -0.041\\
	800   &  & 0.017 & -0.006 & -0.003 & 0.002 & -0.028 & -0.013  &  & 0.021 & -0.006 & -0.003 & 0.004 & -0.023 & -0.010\\
	1600   &  & 0.009 & -0.001 & -0.001 & 0.000 & -0.014 & -0.006 & & -0.001 & -0.008 & 0.002 & 0.003 & -0.008 & -0.008\\
	3200  &  & -0.001 & -0.004 & -0.001 & 0.001 & -0.005 & -0.003 && 0.010 &  0.000 & -0.003 & 0.000 & -0.006 & -0.002\\
	& & \multicolumn{13}{c}{Standard Deviation / Standard Error}\\
	200  &  &1.361 & 1.129 & 1.338 & 1.142 & 1.452 & 1.179 &  & 1.365 & 1.234 & 1.520 & 1.168 & 1.484 & 1.157\\
	800  &  & 1.049 & 0.954 & 1.031 & 1.022 & 1.040 & 0.984 &  &1.035 & 0.979 & 1.036 & 1.010 & 1.033 & 0.990\\
	1600  &  & 1.054 & 1.031 & 1.022 & 1.008 & 1.024 & 1.006 &  &0.998 & 0.966 & 1.005 & 0.972 & 1.016 & 0.975\\
	3200  &  & 1.031 & 0.973 & 0.974 & 0.997 & 1.040 & 0.992 &  & 1.045 & 1.041 & 1.015 & 0.948 & 0.976 & 1.014\\
	&  & \multicolumn{6}{c}{$\rho^{\ast
		}=0 $ $ ,  $ $\omega^{\ast}=0.65$} &  & \multicolumn{6}{c}{$\rho^{\ast
		}=0.65 $ $ ,  $ $\omega^{\ast}=0.65$}\\\cline{4-7}\cline{11-14}
	&  & \multicolumn{13}{c}{Bias}\\
	200 &  & -0.190 & -0.262 & 0.228 & 0.254 & -0.171 & -0.120 &  & -0.205 & -0.280 & 0.225 & 0.261 & -0.162 & -0.116 \\
	800   &  & -0.226 & -0.241 & 0.238 & 0.238 & -0.098 & -0.090 &  & -0.233 & -0.243 & 0.235 & 0.240 & -0.096 & -0.090 \\
	1600   &  &-0.231 & -0.238 & 0.236 & 0.235 & -0.089 & -0.084 &  & -0.227 & -0.237 & 0.233 & 0.235 & -0.089 & -0.084 \\
	3200  &  & -0.235 & -0.237 & 0.236 & 0.235 & -0.083 & -0.082 &  & -0.230 & -0.238 & 0.234 & 0.236 & -0.085 & -0.082 \\
	&  & \multicolumn{13}{c}{Standard Deviation / Standard Error}\\
	200  &  & 1.183 & 1.151 & 1.328 & 1.129 & 1.328 & 1.075 &  & 1.182 & 1.189 & 1.322 & 1.111 & 1.317 & 1.183 \\
	800  &  & 0.812 & 0.888 & 1.008 & 0.819 & 0.826 & 0.861 &  & 0.979 & 1.012 & 1.035 & 0.991 & 1.038 & 1.025 \\
	1600  &  & 0.997 & 1.042 & 0.998 & 0.988 & 1.000 & 0.975 &  & 0.989 & 1.071 & 0.982 & 0.965 & 1.011 & 1.003 \\
	3200  &  & 1.021 & 1.080 & 1.002 & 0.995 & 0.971 & 1.022 &  & 1.084 & 1.162 & 1.024 & 1.030 & 1.018 & 1.032\\
	\\\hline
\end{tabular}
\end{sidewaystable}

As pointed out in Section~\ref{discuss}, another situation in which ignoring
Markov dependence of the regimes is costly involves outcome equations that
contain autoregressive dynamics. To demonstrate numerically the difficulties
in such a case, 1000 artificial samples of various sizes are generated
according to the Markov-switching autoregression
\begin{equation}
Y_{t}=\mu_{1}^{\ast}(S_{t})+\phi^{\ast}Y_{t-1}+\sigma_{1}^{\ast}(S_{t}%
)U_{1,t}, \label{ardgp}%
\end{equation}
with $\phi^{\ast}=0.9$; the remaining parameter values and the generating
mechanisms of $\{Z_{t}\}$, $\{S_{t}\}$ and $\{(U_{1,t},U_{2,t})\}$ are the
same as in earlier simulation experiments. For each artificial sample, the
parameters of the regime-switching autoregressive model
\begin{equation}
Y_{t}=\mu(S_{t})+\phi Y_{t-1}+\sigma(S_{t})\varepsilon_{t}, \label{armodel}%
\end{equation}
are estimated by maximizing the quasi-log-likelihood function associated with
it under the assumption that the regime variables $\{S_{t}\}$ are i.i.d., with
$\Pr(S_{t}=1)=\bar{\vartheta}$, and the noise variables $\{\varepsilon_{t}\}$
are i.i.d., independent of $\{S_{t}\}$, with $\varepsilon_{t}\sim
\mathcal{N}(0,1)$.

The Monte Carlo results reported in Table~\ref{MC-t2} reveal substantial
finite-sample bias in the case of the QML estimators of the intercepts
$\mu(1)$ and $\mu(2)$. The QML estimators of $\sigma(1)$, $\sigma(2)$ and
$\phi$ generally exhibit little bias, which may be partly due to the fact that
the simulation design is such that the values of $\phi^{\ast}$ and $\sigma
_{1}^{\ast}$ are the same regardless of the realized regime. Unlike the HMM
case considered before, estimated standard errors are not always accurate as
approximations to the finite-sample standard deviation of the QML estimators
in the autoregressive model, even for a parameter such as $\sigma(1)$, which
is estimated with little bias. We note that qualitatively similar results are
obtained when, in addition to $Y_{t-1}$, an exogenous covariate $W_{t}$,
generated as in the previous experiments, is included in the right-hand sides
of (\ref{ardgp}) and (\ref{armodel}). {\footnotesize \begin{table}[ptb]
\caption{Bias and Standard Deviation of QML Estimators (Markov-Switching
Autoregressive Model)}%
\label{MC-t2}
\begin{center}
{\footnotesize
\begin{tabular}
[c]{lcccccccccccc}\hline\hline
$T$ &  & $\mu(1)$ & $\mu(2)$ & $\sigma(1)$ & $\sigma(2)$ & $\phi$ &  &
$\mu(1)$ & $\mu(2)$ & $\sigma(1)$ & $\sigma(2)$ & $\phi$\\\hline
&  & \multicolumn{5}{c}{$\rho^{\ast}=0$} &  & \multicolumn{5}{c}{$\rho^{\ast
}=0.8$}\\\cline{3-7}\cline{9-13}
&  & \multicolumn{11}{c}{Bias}\\
200 &  & -0.500 & 0.222 & -0.093 & -0.141 & -0.012 &  & -0.765 & 0.468 &
-0.114 & -0.135 & -0.049\\
800 &  & -0.400 & 0.039 & 0.042 & -0.067 & 0.002 &  & -0.626 & 0.288 & 0.027 &
-0.060 & -0.034\\
1600 &  & -0.440 & 0.023 & 0.086 & -0.049 & 0.004 &  & -0.753 & 0.276 &
0.096 & -0.049 & -0.031\\
3200 &  & -0.462 & 0.013 & 0.115 & -0.036 & 0.005 &  & -0.699 & 0.249 &
0.103 & -0.035 & -0.028\\
&  & \multicolumn{11}{c}{Standard Deviation / Standard Error}\\
200 &  & 1.546 & 1.214 & 1.764 & 1.563 & 1.140 &  & 0.438 & 1.261 & 0.542 &
1.008 & 1.023\\
800 &  & 1.282 & 0.907 & 1.335 & 1.177 & 0.997 &  & 1.228 & 0.928 & 1.347 &
1.121 & 0.954\\
1600 &  & 1.396 & 0.916 & 1.335 & 0.708 & 0.988 &  & 1.542 & 1.011 & 1.504 &
1.066 & 1.008\\
3200 &  & 1.424 & 0.897 & 1.398 & 0.913 & 0.968 &  & 1.078 & 0.803 & 1.102 &
0.775 & 0.976\\\hline
\end{tabular}
}
\par
{\footnotesize  }
\end{center}
\end{table}}

\section{Conclusion}

\label{conclusion}

In this paper, we have considered QML estimation of the parameters of a
generalized HMM with exogenous covariates and a finite hidden state space. A
distinguishing feature of our approach is that it allows the regime sequence
to be a temporally inhomogeneous Markov chain with covariate-dependent
transition probabilities. It has been shown that a mixture model with
independent regimes is robust in the presence of correlated Markov regimes, in
the sense that the parameters of the outcome equation can be estimated
consistently by maximizing the quasi-likelihood function associated with the
misspecified mixture model.

One possible application of our main result is to exploit it to construct
tests for the number of regimes in HMMs with covariate-dependent transition
probabilities, adopting a QML-based approach analogous to that of
\cite{chowhite07}. As is well known, such testing problems are non-standard
and typically involve unidentifiable nuisance parameters, parameters that lie
on the boundary of the parameter space, singularity of the information matrix,
and non-quadratic approximations to the log-likelihood function.

\bibliographystyle{chicago}
\bibliography{markov1}

\end{document}